\documentclass[aps,orb,twocolumn,amssymb,showpacs]{revtex4}

\usepackage[dvipdf]{epsfig}

\usepackage{color}

\def\be{\begin{equation}}

\def\ee#1{\label{#1}\end{equation}}

\usepackage{graphicx}

\usepackage{amsfonts}

\usepackage{amsmath,amssymb}

\usepackage{amsbsy}

\begin{document}

\title{Fokker-Planck type equations for a simple gas and for a semi-relativistic 
 Brownian motion from a relativistic kinetic theory}

\author{Guillermo Chac\'{o}n--Acosta} \email{gca@xanum.uam.mx}
\affiliation{Departamento de F\'\i sica, Universidad Aut\'onoma Metropolitana--Iztapalapa,  M\'exico D. F.
09340, M\'exico  }

\author{Gilberto M. Kremer} \email{kremer@fisica.ufpr.br}
\affiliation{Departamento de F\'\i sica, Universidade Federal do Paran\'a,  81531-990 Curitiba, Brazil}

\begin{abstract}

A covariant Fokker-Planck type equation for a simple gas and an equation for the Brownian motion are derived
from a relativistic kinetic theory based on the Boltzmann equation. For the simple gas the dynamic friction
four-vector and the diffusion tensor are identified and written in terms of integrals which take into account
the collision processes. In the case of Brownian motion, the Brownian particles are considered as
non-relativistic whereas the background gas behaves as a relativistic gas. A general expression
for the semi-relativistic viscous friction coefficient is obtained and the particular case of constant
differential cross-section is analyzed for which the non-relativistic and ultra relativistic limiting cases
are calculated.

\end{abstract}

\pacs{51.10.+y; 05.40.Jc }

\maketitle


\section{Introduction}

 The problem of Brownian motion played a fundamental role in the
early verifications of kinetic theory after the estimation of the Avogadro number by Perrin in 1908  on the
basis of Einstein's theory. Brownian motion and diffusion processes have a wide range of applications that goes
from chemistry, solid state, quantum physics, plasmas, astronomy and astrophysics to social sciences, life
sciences and biology~\cite{Han100}.

 Other kinds of applications can be found if we extend the Brownian motion to the regime of the theory
of relativity. The relativistic Brownian motion could have important applications in plasma physics, in high
energy physics \cite{Hees}, in astrophysics -- for example in the analysis of the gamma ray burst jets where a
Brownian motion of the electrons in the electrostatic wave field could be present \cite{Dk}  -- in relativistic
corrections to the Sunyaev-Zeldovich effect \cite{Itho}, and also in the evolution of dark matter \cite{Bersch}.

 Recently, several different approaches of a relativistic theory of stochastic processes have been used for the
implementation of the Brownian motion concept into the theory of special relativity \cite{RZ,Debb1,Debb2}, some
of them generalized their results to a curved space-time \cite{Debb3}, and studied the problem
from the point of view of reaching quantum theory with Nelson's method
\cite{OH}, and also from quantum gravity \cite{Hu,Hu2}. A comparison among different relativistic stochastic
processes existing in the literature can be found in \cite{Debb4}.

 In the works \cite{Hang1,Hang2}, Dunkel and H\"anggi constructed a relativistic theory for the Brownian
motion starting with a relativistic version of the Langevin equation and obtained that the related Fokker-Planck
equation can be written as a continuity equation, but with different expressions for the flux density by using
different interpretations of the stochastic processes. Moreover, they found that only one of the approaches
leads to the equilibrium relativistic Maxwell-J\"{u}ttner distribution function, while the others differ through
factors that depend on the energy of the particles. In a recent paper by the same authors~\cite{Hang3} they have
started from a microscopic collision model and constructed a criteria to identify the equilibrium distribution
of the particles. Firstly they have used this criteria in the non-relativistic case and recognized the
Maxwellian distribution function, then they applied the same criteria in the relativistic case and, following
their previous results, found that the relativistic distribution function that satisfies their criteria, differs
from the relativistic Maxwell-J\"{u}ttner distribution function by a factor proportional to the inverse of the
relativistic kinetic energy. This result does not agree with the one which comes out from the relativistic
kinetic theory based on the Boltzmann equation where the only distribution function which implies a vanishing
collision term in equilibrium is the Maxwell-J\"{u}ttner distribution function (see e.g.
~\cite{sygrg}, ~\cite{K&C} and ~\cite{dG}).

 It is important to have a theory of Brownian motion founded on kinetic theory, since that
theory could give the connection between the fundamental microscopic dynamics and macroscopic schemes which
have measurable properties. With such a theory one can obtain the equations that govern the motion, and also
calculate the transport coefficients as a function of the properties of the specific system. Is the aim of the
present work to investigate a relativistic generalization of the Fokker-Planck equation and of the equation for
the Brownian motion within the framework of the relativistic kinetic theory based on the Boltzmann equation.

 This work is structured as follows. In Sec. II a covariant version of the  Fokker-Planck equation is derived
from a relativistic kinetic theory based on the Boltzmann equation and applied to a system with only one
constituent. The same assumptions of grazing collisions  valid for a non-relativistic gas (see e.g. \cite{Cer}
and \cite{ChCw}) are also considered here, which consist in small deflections on the scattering angle and small
changes in the momentum of the particles at collision. The dynamic friction four-vector and the diffusion tensor
are identified and written in terms of integrals which take into account the collision processes. In Sec. III a
mixture of two species is considered where one of the components has a small particle number density and whose
particles have a large mass when compared with the other component. Those conditions allow us to call this case
a Brownian motion by analogy with the fluctuation process (see e.g.~\cite{Gard}). As in the non-relativistic
case (see e.g.~\cite{WChU}) the former is identified with the Brownian particles and the latter
with the background gas. The Brownian particles are considered as non-relativistic whereas the background gas
behaves as a relativistic gas which may alternate from the non-relativistic limit to the ultra-relativistic
limit depending on the  ratio between the rest energy of the particles of the gas and of the thermal energy of
the mixture. A general expression for the semi-relativistic viscous friction coefficient is obtained and  the
particular case of constant differential cross-section is analyzed for which the non-relativistic and ultra
relativistic limiting cases are calculated. Concluding remarks are given in Sec. IV.


\section{Relativistic Fokker-Planck  equation for a simple gas}

 Let  a  particle of a simple relativistic gas be characterized by  its rest mass $m$ and by the space-time
coordinates $(x^\alpha)=(ct, {\bf x})$ and   momentum four-vector $(p^\alpha)=(p^0, {\bf p})$, where the
component of the momentum four-vector $p^0$ is constrained by $p^0=\sqrt{\vert {\bf p}\vert +m^2c^2}$. The
state of the relativistic gas in the phase space is characterized by the one-particle distribution function
$f(x^\alpha, p^\alpha)= f({\bf x}, {\bf p}, t)$, such that $ f({\bf x, p},t)d^3{ x}\, d^3{ p}$ gives at time $t$
the number of particles in the volume element $d^3x$ about ${\bf x}$ and with momenta in a range $d^3p$ about
$\bf p$.

 For an elastic collision of two particles with momentum four-vectors denoted by
$p^{\alpha}$ and $p^{\alpha}_*$, the energy-momentum conservation law reads $
p^{\alpha}+p^{\alpha}_*=p^{\prime\alpha} +p^{\prime\alpha}_*,$ where the quantities $p^{\prime\alpha}$,
$p^{\prime\alpha}_*$ denote the values of the momentum four-vectors after  collision.

 The one-particle distribution function satisfies the re\-lativistic Boltzmann
equation (see e.g~\cite{K&C,dG})
 \be
 p^{\alpha}{\partial f\over \partial x^\alpha} +m{\partial
 fK^{\alpha}\over \partial p^\alpha} =\int \left(f_*'f'-f_*f\right)
 \,F\,\sigma\,d\Omega {d^3p_*\over p_{*0}},
 \ee{1}
where the usual abbreviations $f_*'\equiv f({\bf x,p}_*',t),$ $f'\equiv f({\bf x,p'},t),$ $f_* \equiv f({\bf
x,p}_*,t),$ and $f \equiv f({\bf x,p},t)$ were introduced. In the above equation, $\sigma$ denotes a
differential cross-section,  $d\Omega$ an element of solid angle which characterizes the scattering process,
$K^\alpha$ the Minkowski external force and $F=\sqrt{(p^\alpha_*p_\alpha)^2-m^4c^4}$ the so-called invariant
flux.

Under the assumption of grazing collisions that could take place in long range interactions, only small changes
in the momentum of the particles occur due to small deflections in scattering angle. Then the collision term in
the right-hand side of the Boltzmann equation (\ref{1}), denoted by ${\cal Q}(f,f_\ast)$,  can be approximated
as follows.

 First new variables are introduced, namely,  the total $P^\alpha$ and the relative $Q^\alpha$
momentum four-vectors, defined by
  \be
  P^\alpha=p^\alpha+p^\alpha_*=P^{\prime\alpha},\quad
  Q^\alpha=p^\alpha-p^\alpha_*,\quad
  Q^{\prime\alpha}=p^{\prime\alpha}-p^{\prime\alpha}_*.
  \ee{2}

 For these quantities the following relationships hold $P^\alpha Q_\alpha=0$ and
$P^2=Q^2+4m^2c^2$ where $P^\alpha P_\alpha=P^2$ and $Q^\alpha Q_\alpha=-Q^2$.

 Further, from the energy-momentum conservation law and eq. (\ref{2})
one can write the differences between the post- and pre-collision momentum four-vectors as
 \be
  \begin{cases}
  \Delta p^\alpha\equiv p^{\prime\alpha}-p^\alpha={1\over2}\left(
  Q^{\prime\alpha}-Q^\alpha\right)\equiv {1\over2}\Delta Q^\alpha,\cr
  \Delta p^\alpha_*\equiv p_*^{\prime\alpha}-p_*^\alpha=-{1\over2}\left(
  Q^{\prime\alpha}-Q^\alpha\right)\equiv -{1\over2}\Delta Q^\alpha.
  \end{cases}
 \ee{3}

 Hence for small changes of the momentum of the particles at
collision one can expand the one-particle distribution function in Taylor series, which up to the second-order
terms reads
 \be
  f(p^{\prime i})\approx f(p^i) +{1\over2}\Delta Q^i \frac{\partial
  f}{\partial p^i}+\frac{1}{8}\Delta Q^i \Delta Q^j \frac{\partial^2
  f}{\partial p^i \partial p^j},
 \ee{4}
with a similar expression for $f(p^{\prime i}_*)$, by making the
changes  $p^i\rightarrow p^i_*$ and $\Delta Q^i\rightarrow -\Delta
Q^i$. In these expressions we take only the first two derivatives
because the other contributions are of smaller order thanks to the
hypothesis of grazing collisions.

 Now it is possible to approximate the collision term  of the
Boltzmann equation (\ref{1}) as
  \[
  {\cal Q}(f,f_\ast)=\int\left[\Delta Q^i\frac{\partial}{\partial Q^{i}} (f f_*) \right.
  \]
\be\left.+\frac{1}{2} \Delta Q^i
 \Delta Q^j\frac{\partial}{\partial Q^{i}}\frac{\partial}{\partial
 Q^{j}}(f f_*)\right]\,F\,\sigma\,d\Omega {d^3p_*\over p_{*0}},
\ee{5}
with the help of the relationship
 \be
 \frac{\partial}{\partial Q^{i}} = \frac{1}{2} \left( \frac{\partial}{\partial p^{i}_*}
 - \frac{\partial}{\partial p^{i}} \right).
\ee{6}

In order to transform the integral (\ref{5}),  the center-of-mass system is chosen where the spatial components
of the total momentum four-vector vanish, i.e.
 \be
 (P^{\alpha})=(P^0,{\bf 0}),
 \qquad\qquad (Q^{\alpha})=(0,{\bf Q}).
 \ee{7}
Now  the element of solid angle in (\ref{5}) can be written as $d\Omega=\sin\Theta d\Theta d\Phi$, where
$\Theta$ and $\Phi$ are polar angles of $Q^{\prime\alpha}$ with respect to $Q^{\alpha}$ and such that $\Theta$
represents the scattering angle. Further,  without loss of generality,  $Q^{\alpha}$ is chosen in the direction
of the 3-axis, so that one can write $Q^{\alpha}$ and $Q^{\prime\alpha}$ as:
 \be
 (Q^{\alpha})=Q
 {\begin{pmatrix}0\cr
 0\cr 0\cr 1\end{pmatrix}}, \qquad\qquad (Q^{\prime\alpha})=Q{\begin{pmatrix}0\cr
 \sin\Theta\cos\Phi\cr \sin\Theta\sin\Phi\cr \cos\Theta\end{pmatrix}}.
 \ee{8}

 By using the above representations it is easy to calculate the following integrals in the variable $0\leq
\Phi\leq 2\pi$, yielding
 \be
 \int \Delta Q^i {F\over p_{*0}}\sigma \sin\Theta\, d\Theta\, d\Phi= -
 Q^i Q\Sigma,
 \ee{9}
 \be
  \int\Delta Q^i \Delta Q^j {F\over p_{*0}}\sigma \sin\Theta\, d\Theta\, d\Phi =
  -(Q^2\eta^{ij}+Q^iQ^j)Q\Sigma.
 \ee{10}
 Note that that the differential cross-section is a function of $\sigma=\sigma(Q,\Theta)$
 whereas the invariant flux is given by $F/p_{*0}=Q$.
 Above $\Sigma$ denotes the following integral
 \be
 \Sigma =2\pi \int( 1-\cos\Theta)  \sigma  \sin\Theta \,d\Theta,
 \ee{11a}
 and $\eta^{ij}$ are the spatial components of the metric tensor
 $(\eta_{\alpha\beta})={\rm diag}(1,-1,-1,-1)$.
 In order to obtain the integral (\ref{10}) the following
 approximation was taken into account  $\sin^2\Theta \approx 2(1-\cos\Theta)$, since
 only grazing collisions between the particles with small scattering
 angles are considered.

  By differentiating (\ref{10}) with respect to $Q^j$ and considering
 the relationship $\partial Q/\partial Q^j=-Q_j/Q$, one can obtain  the following connection
 between the  integrals (\ref{9}) and (\ref{10}):
 \be
 \frac{\partial}{\partial Q^j}\int \Delta Q^i \Delta Q^j Q\sigma d\Omega= 2 \int \Delta Q^i
 Q\sigma d\Omega.
 \ee{11}

 Now by making use of  equations (\ref{6}) and (\ref{11}), the collision integral (\ref{5}) becomes
 \[
 {\cal Q}(f,f_\ast)=\frac{1}{4} \int \frac{\partial}{\partial p^{i}_*} \left\{\int \frac{\partial (ff_*)}{\partial
Q^j}\Delta Q^i \Delta Q^j Q \sigma d\Omega \right\}d^3p_*
 \]
 \be
 -\frac{1}{4} \frac{\partial}{\partial p^{i}}
\left\{\int\frac{\partial (ff_*)}{\partial Q^j}\Delta Q^i \Delta Q^jQ \sigma d\Omega d^3p_*\right\}.
 \ee{12}
 The first term on the
 right-hand side of the above equation vanishes, since the hypothesis of grazing collisions
 is used and it is
 possible  to convert -- thanks to the divergence theorem --  the volume integral in the
 momentum space into an integral at an infinitely far surface where the distribution functions tend to
 zero.  The second term on the right-hand side can be manipulated by
 using equations   (\ref{6}) and (\ref{11}), yielding
  \[
 {\cal Q}(f,f_\ast)=\frac{1}{4} \frac{\partial}{\partial p^{i}}\left\{
 2 \int ff_*\Delta Q^i Q
\sigma d\Omega d^3p_*
 \right.
 \]
 \be
 \left.
-{1\over2}\int \left(\frac{\partial}{\partial p^{j}_*}-\frac{\partial}{\partial p^{j}} \right)
  (ff_*\Delta Q^i \Delta Q^j Q \sigma d\Omega) d^3p_*
 \right\}.
 \ee{13a}
 By invoking the divergence theorem again and using (\ref{6}), it
 follows that
 \be
 {\cal Q}(f,f_\ast)=-{\partial\over \partial
 p^i}\left[fA^i-{\partial fD^{ij}\over\partial p^j}\right],
 \ee{13}
 where the spatial components of the coefficient of dynamic friction $A^i$ and  the diffusion
 coefficient $D^{ij}$ are given by
 \be
 A^i=\int f_*\Delta p^i_*\,F \sigma d\Omega {d^3p_*\over p_{*0}},
 \ee{14}
 \be
 D^{ij}={1\over2}\int f_*\Delta p^i_*\Delta p^j_*\,F \sigma d\Omega {d^3p_*\over
 p_{*0}}.
 \ee{15}

 It is important to call attention to the fact that left-hand side of the Boltzmann equation is a scalar
 invariant, but the collision term written in the expression (\ref{13}) is not
 an invariant. In order to write a covariant form of the
 Boltzmann equation one has to recall that the above results were
 obtained by considering the center-of-mass system. In this system $\Delta
 p^0_*=0$, and one can include the zeroth components
\be
 A^0=\int f_*\Delta p^0_*\,F \sigma d\Omega {d^3p_*\over p_{*0}},
 \ee{14a}
 \be
 D^{i0}=D^{0i}={1\over2}\int f_*\Delta p^i_*\Delta p^0_*\,F \sigma d\Omega {d^3p_*\over
 p_{*0}},
 \ee{15a}
 \be
 D^{00}={1\over2}\int f_*\Delta p^0_*\Delta p^0_*\,F \sigma d\Omega {d^3p_*\over
 p_{*0}},
 \ee{15b}
 into the collision term (\ref{13}) so that
 it becomes a scalar invariant. While the components of the coefficient of dynamic friction $A^i$ and  of the diffusion
 coefficient $D^{ij}$ are related with the changes of the momenta $\Delta
 p^i_*$ at collision, the components $A^0$ and $D^{00}$ take into account
 the energy changes  $\Delta p^0_*$ and  $D^{0i}$ the changes of both $\Delta p^0_*$ and $\Delta p^i_*$.
  They are the temporal components of the four-vector
 $A^\alpha$ and of the four-tensor $D^{\alpha\beta}$.

 Hence the relativistic Boltzmann
 equation (\ref{1}) reduces to the relativistic  Fokker-Planck equation, namely,
 \be
 p^{\alpha}{\partial f\over \partial x^\alpha} +m{\partial
 fK^{\alpha}\over \partial p^\alpha} =-{\partial\over \partial
 p^\alpha}\left[fA^\alpha-{\partial fD^{\alpha\beta}\over\partial p^\beta}\right].
 \ee{16}

 For a spatially homogeneous case without external forces, the Fokker-Planck equation
 (\ref{16}) reduces to
 \be
 {p^0\over c}{\partial f\over \partial t}+{\partial {\cal
 F}^\alpha\over \partial p^\alpha}=0,
 \ee{17}
 which represents a continuity equation in  momentum space with the particle flux density
  ${\cal F}^\alpha$  given by
 \be
 {\cal F}^\alpha=f\left(A^\alpha- {1\over f}{\partial fD^{\alpha\beta}\over\partial p^\beta}\right).
 \ee{18}

The stationary solution of eq. (\ref{17}) for the distribution
function $f$ is obtained by assuming that  the particle flux density
vanishes, i.e. $\mathcal{F}^{\alpha}=0$. This condition is
equivalent to the one that in equilibrium the collision term of the
Boltzmann equation vanishes and the distribution function is
characterized by the Maxwell-J\"{u}ttner distribution \cite{K&C,dG},
namely,
 \be
 f \propto \exp{\left[-\frac{U_{\alpha}p^{\alpha}}{kT}\right]},
 \ee{18a}
 where $U_{\alpha}$ --  such that $U^\alpha U_\alpha=c^2$ -- is the hydrodynamical four-velocity of the gas.

If we insert (\ref{18a}) into  (\ref{18}), by considering $\mathcal{F}^{\alpha}=0$, it follows that
 \be
 kT \tilde A^{\alpha} = - U_{\beta}D^{\alpha
 \beta},\quad \textrm{where}\qquad \tilde A^\alpha=A^{\alpha} - \frac{\partial D^{\alpha \beta}}{\partial p^{\beta}}.
 \ee{18b}
 The representations of the four-vector $\tilde A^\alpha$ and of the four-tensor $D^{\alpha\beta}$
 in terms of the four-momentum $p^\alpha$ read
 \be
 \tilde A^\alpha= A\,p^\alpha,\qquad D^{\alpha\beta}=-\tilde D\,\eta^{\alpha\beta}-D{p^\alpha p^\beta\over
 m^2c^2},
 \ee{18c}
 where $A, \tilde D$ and $D$ are scalar coefficients and
 $D^\alpha_\alpha=-4\tilde D-D$.

 The insertion of the
 representations (\ref{18c}) into the equation (\ref{18b})$_1$  leads to
 \be
 \tilde D=0,\qquad kT\,A={p^\alpha U_\alpha\over m^2c^2}D,
 \ee{18d}
 i.e., only one among the three coefficients is linearly independent.

 Hence, thanks to (\ref{18c}) and (\ref{18d}), the spatially
 homogeneous  Fokker-Planck equation
 (\ref{17}) reduces to
 \be
 {p^0\over c}{\partial f\over \partial t}+{\partial \over \partial p^\alpha}\left[
 {Dp^\alpha\over m^2c^2}\left(f{p^\alpha U_\alpha\over kT}+p^\beta{\partial f\over\partial
 p^\beta}\right)\right]=0.
 \ee{18e}

In a Lorentz rest frame -- where $U^\mu=(c,{\bf0})$ -- one can
obtain from equation
 (\ref{18d})$_2$
 \be
  \frac{mkTA}{D}=\sqrt{1+{\vert{\bf p}\vert^2\over m^2 c^2}}  \approx 1 + \frac{v^2}{2c^2} + \ldots,
 \ee{18f}
 where the leading term, for $v\ll c$, corresponds to the usual relation between
the diffusion coefficient and the friction coefficient.


\section{Relativistic equation for Brownian Motion}

 In this section a mixture of two constituents is considered, where one of the components consist of heavy
particles of rest mass $m_{\rm b}$ while the other by light particles of rest mass $m_{\rm g}$, so that $m_{\rm
b}\gg m_{\rm g}$. The component with light particles describes a rarefied gas with particle number density much
larger than that of the constituent with heavy particles and which characterizes the Brownian particles $n_b \ll
n_g$. The gas is supposed to be at equilibrium with a Maxwell-J\"uttner distribution function
 \be
 f_{\rm g}=f_{\rm g}^{(0)}=\frac{n_{\rm g}}{4 \pi m_{\rm g}^2 ckTK_2(\zeta_{\rm
 g})}\, e^{-\frac{U_{\alpha}p^{\alpha}_{\rm g}}{kT}}.
 \ee{19}
 Above, $n_{\rm g}$ denotes the particle number density of the gas,
 $c$ the speed of light, $k$ the Boltzmann constant, whereas $T$ and $U^\alpha$ are
 the temperature and the four-velocity of the mixture, respectively.
 The symbol $K_2(\zeta_{\rm g})$ refers to a modified Bessel
 function of second kind and $\zeta_{\rm g}=m_{\rm g}c^2/kT$ gives
 the ratio between the rest and thermal energies of the gas particles.

  The two assumptions above for the Brownian constituent -- that it
  has a small particle number density and particles with large mass
  with respect to the gas constituent -- imply that it
  may be considered as a non-relativistic gas with negligible collision
  term with respect to its particles. Hence the Boltzmann equation
  for the one-particle distribution function of the Brownian particles can be written as
  \be
 \frac{\partial f_{\rm b}}{\partial t} +v^i_{\rm b} \frac{\partial f_{\rm b}}{\partial x^i} +
 \frac{\partial(f_{\rm b} F^i)}{\partial
    p_{\rm b}^i} = \int (f_{\rm g}^{(0)\prime} f^\prime_{\rm b}-f_{\rm g}^{(0)}f_{\rm b})
    g_{\o}\sigma d\Omega d^3p_{\rm g}.
  \ee{20}
 The above equation follows easily from the Boltzmann equation (\ref{1}) written for
 a Brownian particle through its multiplication by $c/p^0_{\rm b}$, the introduction of the
 velocity $v^i_{\rm b}=cp^i_{\rm b}/p^0_{\rm b}$ and of the M{\o}ller velocity defined by
  \be
   g_{\o}={cF\over p^0_{\rm b}p^0_{\rm g}}=\sqrt{(\mathbf{v}_{\rm g}
   -\mathbf{v}_{\rm b})^2-\frac{1}{c^2}(\mathbf{v}_{\rm g}
   \times \mathbf{v}_{\rm b})^2}.
   \ee{21}

Since the collisions of the gas particles with the Brownian particles affect very little the depart from
equilibrium of the latter, one can suppose that the distribution function of the Brownian particles can be
written as
 \be
 f_{\rm b} =  f_{\rm b}^{(0)}
 h(p_{\rm b}^i), \qquad f_{\rm b}^{(0)}= e^{-\frac{U_{\alpha}p^{\alpha}_{\rm b}}{kT}}
 \ee{22}
 where $f_{\rm b}^{(0)}$ refers to the exponential factor of a
 Maxwell-J\"uttner distribution function for the Brownian particles
 and $h(p_{\rm b}^i)$ represents a deviation from this
 distribution when we assume space homogeneity. Moreover, based on this assumption one can expand the
 deviation for the post-collision momentum $h(p_{\rm b}^{\prime i})$
 in Taylor series and by keeping up to second order terms in the difference
 $\Delta p_{\rm b}^{ i}=p_{\rm b}^{\prime i}-p_{\rm b}^{ i}$, yields
 \be
 h(p_{\rm b}^{\prime i})=h(p_{\rm b}^{i}) + \Delta p^i_{\rm b} \frac{\partial h}{\partial p_{\rm b}^i}
 +\frac{1}{2}\Delta p^i_{\rm b} \Delta p^j_{\rm b}
\frac{\partial^2 h}{\partial p^i_{\rm b} \partial p^j_{\rm b}}.
 \ee{23}

 Equations (\ref{22}) and (\ref{23}) are introduced into the collision term
 of the Boltzmann equation (\ref{20}) so that it can be written as
 \be
 {\cal Q}(f_{\rm b},f_{\rm g}) =f_{\rm b}^{(0)}{\cal A}
  {\cal I},
 \ee{24}
 where the relationship $ f_{\rm b}^{\prime(0)} f_{\rm
g}^{\prime(0)}= f_{\rm b}^{(0)} f_{\rm g}^{(0)}$ was used. Above ${\cal A}={n_{\rm g}}/{4 \pi m_{\rm g}^2
ckTK_2(\zeta_{\rm g})}$ and ${\cal I}$ is the following integral
 \be
  {\cal I}\!=\!\!\int\!\Biggl[ \Delta p^i_{\rm b} \frac{\partial h}{\partial p_{\rm b}^i}
 +\frac{1}{2}\Delta p^i_{\rm b} \Delta p^j_{\rm b}
\frac{\partial^2 h}{\partial p^i_{\rm b} \partial p^j_{\rm b}}\Biggr]e^{-\frac{U_{\alpha}p^{\alpha}_{\rm
g}}{kT}}g_{\o}\sigma d\Omega d^3p_{\rm g}.
 \ee{25}

The integral (\ref{25}) can be transformed as follows. First one can note that for a non-relativistic Brownian
particle $p_{\rm b}^0=p_{\rm b}^{\prime0}$ so that energy conservation law leads to $p_{\rm g}^0=p_{\rm
g}^{\prime0}$. Further one can introduce a relative velocity  defined by
 \be
 g^i = \frac{cp^i_{\rm g}}{p^0_{\rm g}}-\frac{cp^i_{\rm b}}{p^0_{\rm b}},\qquad
 g^{\prime i} = \frac{cp^{\prime i}_{\rm g}}{p^0_{\rm g}}-\frac{cp^{\prime i}_{\rm b}}{p^0_{\rm
 b}},
 \ee{26}
 such that the difference $\Delta g^i=g^{\prime i}-g^{i} $ can be
 written, thanks to the momentum conservation law, in terms of the difference $\Delta
 p^i$ as follows
 \be
 \Delta g^i=-{c\over p^0_{\rm g}}\left(1+{p^0_{\rm g}\over p^0_{\rm b}}\right)\Delta
 p^i_{\rm b}.
 \ee{27}

 For a relativistic gas of rest massless particles $\vert p^i_{\rm g}/p^0_{\rm
 g}\vert=1$ whereas $\vert p^i_{\rm b}/p^0_{\rm
 b}\vert=v_b/c$, hence one can approximate the difference given by (\ref{27})
 as
  \be
 \Delta g^i\approx-{c\over p^0_{\rm g}}\Delta
 p^i_{\rm b}.
 \ee{28}
 The above approximation is also valid for a gas with massive
 particles, since $m_{\rm b}\gg m_{\rm g}$ and $v_{\rm g}\neq c$ imply
 that $p^0_{\rm g}/ p^0_{\rm b}\ll1$.

 The integral (\ref{25}) can be written in terms of the
 difference $\Delta g^i$ by using (\ref{28}). In the resulting equation,
 two integrations can be performed, the first one
 can be done  by introducing  the scattering
 angle  $\chi$ and the azimuthal angle $\epsilon$  which are the
 spherical angles of $g^{\prime i}$ with respect to $g^i$.  By writing
 the element of solid angle as $d\Omega=\sin{\chi}d\chi d\epsilon$
 and integrating the differences $\Delta g^i$ and $\Delta g^i\Delta g^j$ with respect to the
 azimuthal angle $0\leq\epsilon\leq2\pi$, yields
\be
  \int^{2\pi}_{0} \Delta g^i d\epsilon = -2\pi\left(1-\cos \chi \right)g^i,
\ee{29}
\[
  \int^{2\pi}_{0} \Delta g^i \Delta g^j d\epsilon = 2\pi \Bigl\{\Bigl[(1-\cos
  \chi)^2
  \]
  \be
  -\frac{1}{2}\sin^2 \chi \Bigr] g^i g^j -\frac{\vert\mathbf{g}\vert^2}{2}\eta^{ij} \sin^2 \chi
  \Bigr\},
 \ee{30}
 where $\vert\mathbf{g}\vert^2=g^kg^k$. For the second integration the
 element $d^3p_{\rm g} =\sin \phi d\phi d\vartheta |\mathbf{p}_{\rm g}|^2d|\mathbf{p}_{\rm g}|$
 is written in terms of the spherical coordinates $(|\mathbf{p}_{\rm
 g}|,\vartheta,\phi)$, where $\vartheta$ and $\phi$ are the
 spherical angles of $p^i_{\rm g}$ with respect to $p^i_{\rm b}$.
 Hence the integrations of $g^i$ and $g^ig^j$ with respect to
 $\vartheta$ become
 \be
  \int^{2\pi}_{0}g^i d\vartheta = 2\pi c\left({\vert\mathbf{p}_{\rm g}\vert\over p^0_{\rm g}}
  {p^0_{\rm b}\over|\mathbf{p}_{\rm b}|}\cos \phi-1 \right)\frac{p^i_{\rm b}}{p^0_{\rm b}},
\ee{31}
\[ \int^{2\pi}_{0}  g^ig^j d\vartheta =
2\pi c^2\Biggl\{ \left[\left({\vert\mathbf{p}_{\rm g}\vert\over p^0_{\rm g}}
  {p^0_{\rm b}\over|\mathbf{p}_{\rm b}|}\cos
 \phi-1\right)^2\Biggr.\right. \]
 \be
 \left.\left.  -\frac{1}{2} {\vert\mathbf{p}_{\rm g}\vert^2\over (p^0_{\rm g})^2}
  {(p^0_{\rm b})^2\over|\mathbf{p}_{\rm b}|^2}\sin^2 \phi \right]
 \frac{p^i_{\rm b} p^j_{\rm b}}{(p_{\rm b}^{0})^2} -\frac{1}{2}\frac{|\mathbf{p}_{\rm g}|^2}
 {(p_{\rm g}^{0})^2}\eta^{ij} \sin^2 \phi  \right\}.
 \ee{32}

  The invariant flux $F=\sqrt{(p_{{\rm g}\alpha}\,p^{\alpha}_{\rm b})^2-m_{\rm b}^2m_{\rm g}^2c^4}$
   can be approximated for the case of a non-relativistic Brownian
   particle by
    \be
  F={p_{\rm g}^0p_{\rm b}^0 g_{\o}\over c}\approx p^0_{\rm b} |\mathbf{p}_{\rm g}| \left(1 -
   \frac{|\mathbf{p}_{\rm b}|}{p_{\rm b}^0}\frac{p^0_{\rm g}}{|\mathbf{p}_{\rm g}|}\cos
   \phi\right),
   \ee{33}
   by considering terms up to the first  order of $|\mathbf{p}_{\rm b}|/p_{\rm b}^0=v_{\rm
   b}/c$. Note that the above approximation is valid if ${p^0_{\rm g}}/{|\mathbf{p}_{\rm
   g}|}$   does not diverge, but in the following we shall show
   that after some transformations such a term disappears from the final form of the
   integral $\cal I$.

   Since the differential cross-section is a function of the
   invariant flux and of the scattering angle it can be approximated by
   \be
   \sigma=\sigma(F,\chi)\approx\sigma(|\mathbf{p}_{\rm g}|,\chi)
   \left( 1- \frac{\partial \ln\sigma}{\partial |\mathbf{p}_{\rm g}|}
   \frac{|\mathbf{p}_{\rm b}|}{p_{\rm b}^0}{p^0_{\rm g}}\cos \phi
   \right).
   \ee{34}

 Now by using  (\ref{28}) through (\ref{34}) in equation  (\ref{25}) and
 integrating the resulting equation with respect to the angle
 $0\leq\phi\leq\pi$ it follows that
 \[
 {\cal I}= 4\pi^2 c \int e^{-\frac{U_{\alpha}p^{\alpha}_{\rm g}}{kT}}
  |\mathbf{p}_{\rm g}|^3 \sigma(|\mathbf{p}_{\rm g}|,\chi)
 (1-\cos \chi)\sin \chi   \]
 \[
\times\left\{ -2p^0_{\rm g}\frac{\partial h}{\partial p^i_{\rm b}} \frac{p^i_{\rm b}}{p^0_{\rm b}} \left[1+
\frac{1}{3}\left( 1+ {|\mathbf{p}_{\rm g}|}\frac{\partial \ln\sigma}{\partial |\mathbf{p}_{\rm g}|}\right)
\right]\right.
 \]
 \be
 \left.-
 \frac{2}{3}(p_{\rm g}^0)^2 \frac{\partial^2 h}{\partial p^i_{\rm b} \partial
 p^j_{\rm b}}\eta^{ij} \frac{|\mathbf{p}_{\rm g}|^2}{(p^{0}_{\rm g})^2}
 \right\}d\chi\frac{d|\mathbf{p}_{\rm g}|}{p^0_{\rm g}}.
 \ee{35}

In order to integrate  by parts the derivative of the cross section
 $\sigma$ with respect to $|\mathbf{p}_{\rm g}|$ a Lorentz rest frame in which
$(U^{\alpha})=(c,\mathbf{0})$ is chosen, so that one can obtain
 \[
    \int  e^{-\frac{cp^{0}_{\rm g}}{kT}}|\mathbf{p}_{\rm g}|^4
    \frac{\partial \sigma}{\partial |\mathbf{p}_{\rm g}|}d|\mathbf{p}_{\rm g}|
    = -\int  e^{-\frac{cp^{0}_{\rm g}}{kT}}\Biggl[
    4|\mathbf{p}_{\rm g}|^3
  \]
 \be
    -\frac{c}{kTp^0_{\rm g}}|\mathbf{p}_{\rm g}|^5\Biggr]
    \sigma(|\mathbf{p}_{\rm g}|,\chi)d|\mathbf{p}_{\rm g}|.
 \ee{36}
 Now the substitution  of the
 above result into (\ref{35}) leads to
  \[
  \mathcal{I}=- \frac{8}{3}\pi^2 c \int e^{-\frac{cp^{0}_{\rm g}}{kT}}
 |\mathbf{p}_{\rm g}|^5\sigma(|\mathbf{p}_{\rm g}|,\chi)(1-\cos \chi)\sin \chi
 \]
 \be
 \times \left[ \frac{cp^i_{\rm b}}{kTp^0_{\rm b}}\frac{\partial h}{\partial p^i_{\rm b}}
 +\eta^{ij} \frac{\partial^2 h}{\partial
 p^i_{\rm b}\partial p^j_{\rm b}} \right]d\chi
 \frac{d|\mathbf{p}_{\rm g}|}{p^0_{\rm g}}.
 \ee{37}

The final form of the  collision term (\ref{24}) in terms of the
distribution function of the Brownian particle can be obtained from
(\ref{37}) by using the representation (\ref{22}), where the
differentiation was also performed in a Lorentz rest reference
frame, namely
 \be
   {\cal Q}(f_{\rm b},f_{\rm g})= \eta \left[m_{\rm b}kT
   \frac{\partial^2 f_{\rm b} }{\partial p^i_{\rm b} \partial p^j_{\rm b}}
   +\frac{\partial f_{\rm b} p^i_{\rm b}}{\partial p^i_{\rm b}}
    \right],
 \ee{38}
 so that the Boltzmann equation for the Brownian particle can be
 written as
 \be
 \frac{\partial f_{\rm b}}{\partial t} +
 \frac{\partial(f_{\rm b} F^i)}{\partial
    p_{\rm b}^i} =\eta \left[m_{\rm b}kT
   \frac{\partial^2 f_{\rm b} }{\partial p^i_{\rm b} \partial p^j_{\rm b}}
   +\frac{\partial f_{\rm b} p^i_{\rm b}}{\partial p^i_{\rm b}}
    \right].
 \ee{38a}
In (\ref{38}) and (\ref{38a}) $\eta$ is the so-called viscous friction coefficient for a semi-relativistic case
and it is given by
 \[
 \eta = \frac{2}{3} \frac{n_{\rm g} \pi}{m_{\rm b}(m_{\rm g}kT)^2K_2(\zeta_{\rm g})}
 \]
\be
  \times   \int \sigma(|\mathbf{p}_{\rm g}|,\chi)(1- \cos \chi)\sin
    \chi d\chi e^{-\frac{cp_{\rm g}^{0}}{kT}}|\mathbf{p}_{\rm g}|^5
    \frac{d|\mathbf{p}_{\rm g}|}{p^0_{\rm g}}.
\ee{39}

Equation (\ref{38a}) is the extension to a background gas of relativistic particles of the Fokker-Planck type
equation for Brownian motion which was analyzed  by Chandrasekhar \cite{Chsk}, Green \cite{Green} and Wang Chang
and Uhlenbeck \cite{WChU} for  the case of a non-relativistic background  gas. The relativistic corrections
appear here in the viscous friction coefficient.

In order to evaluate the integral for the viscous friction coefficient (\ref{39}) it is necessary to know the
differential cross-section $\sigma$. For a constant differential cross-section the integration in the angle
$0\leq\chi\leq\pi$ is straightforward and by introducing a new variable  $y=p^0_{\rm g}/m_{\rm g}c$ so that
$|\mathbf{p}_{\rm g}|=m_{\rm g}c\sqrt{y^2-1}$, the friction viscous coefficient becomes
 \be
    \eta = \frac{4\pi}{3} \frac{n_{\rm g} \sigma}{m_{\rm b}(m_{\rm g}kT)^2K_2(\zeta_{\rm g})}
    \int^{\infty}_1  e^{-\zeta_{\rm g}y}(y^2-1)^{2}dy,
 \ee{41}
which leads through integration to the final form of the viscous friction coefficient of the Brownian particles
in a background gas of relativistic particles:
 \be
    \eta= \frac{32\pi n_{\rm g} \sigma kT}{3m_{\rm b}cK_2(\zeta_{\rm g})} \frac{e^{-\zeta_{\rm g}}}
    {\zeta_{\rm g}^2}\left( 3+3\zeta_{\rm g}+\zeta_{\rm g}^2\right).
 \ee{43}

 For low temperatures the rest energy of the background gas $m_{\rm g}c^2$  is much larger
 than the thermal energy $kT$, so that $\zeta_{\rm g} \gg 1$ and the semi-relativistic viscous friction
 coefficient can be approximated by
 \be
    \eta = \frac{32n_{\rm g} \sigma}{3m_{\rm b}} \sqrt{2m_{\rm g}\pi kT}
    \left[1+ \frac{9}{8 \zeta_{\rm g}}+ \frac{9}{128 \zeta_{\rm g}^2} +
    \ldots\right].
 \ee{44}
By using the hard-sphere cross-section of a non-relativistic  gas $\sigma=d^2/4$ -- where  $d$ denotes the
diameter of a particle -- the first term  in the expression (\ref{44}) reduces to the result obtained by Wang
Chang and Uhlenbeck~\cite{WChU}. The other terms in the series are related with relativistic corrections.

 For very high temperatures the parameter $\zeta_{\rm g} \ll 1$ and
 this condition characterizes the ultra-relativistic regime. In this
 case the approximation for the viscous friction coefficient reads
 \be
    \eta \!=\! \frac{16\pi n_{\rm g}\sigma kT}{m_{\rm b}c}
    \Biggl[1+\frac{\zeta_{\rm g}^2}{12}+\frac{\zeta_{\rm g}^4}{64}
    \Biggl(\!1+4\ln\Biggl(\frac{\zeta_{\rm
    g}}{2}\Biggr)\!+4\gamma\Biggr)\!+
    \ldots\Biggr],
 \ee{45}
where $\gamma=0.577215664\ldots$ is the Euler constant.  In this case the first term is the leading one and the
others are corrections to the former.

\section{Concluding remarks}

In this work we obtain a manifestly covariant relativistic Fokker-Planck type equation for the evolution of the
distribution function of a simple gas under the assumption of grazing collisions. We also get a
semi-relativistic Fokker-Planck equation for the Brownian motion.

Although the Fokker-Planck equation (\ref{16}) and the equation of the Brownian motion (\ref{38a}) are alike --
at least in the non-relativistic limit -- one has to be very careful with the temptation to obtain the latter
from the former. One can write a Fokker-Plack equation like the one given in (\ref{16}) for a mixture of two
constituents where one of constituents has a small particle number density with respect to the other, so that
only  collisions between dissimilar particles are taken into account. However, in Sec. II  only grazing
collisions between the particles were taken into account, while in the case analyzed in Sec. III the collisions
between the particles of the background gas and the Brownian particles were not restricted to grazing
collisions.

The extension of the Fokker-Planck type equation for a simple gas in the presence of a gravitational field is
straightforward, since the Boltzmann equation in gravitational fields is written as (see e.g.~\cite{K&C})
 \be
 p^\mu{\partial f\over
\partial x^\mu}-\Gamma_{\mu\nu}^\sigma p^\mu p^\nu{\partial f\over\partial p^\sigma}=
\int\left(f_*'f'-f_*f\right) \,F\,\sigma\,d\Omega \sqrt{g}{d^3p_*\over p_{*0}}.
 \ee{46}
Here $\sqrt{g}=\sqrt{-\det(g_{\mu\nu})}$ where $g_{\mu\nu}$ denotes the metric tensor and
$\Gamma_{\mu\nu}^\sigma$ is the Cristoffel symbol. Indeed, the Fokker-Planck equation for this case becomes
 \be
 p^\mu{\partial f\over
\partial x^\mu}-\Gamma_{\mu\nu}^\sigma p^\mu p^\nu{\partial f\over\partial p^\sigma}=
-{\partial\over \partial
 p^\mu}\left[fA^\mu-{\partial fD^{\mu\nu}\over\partial
 p^\nu}\right],
 \ee{47}
 where the  dynamic friction four-vector $A^\mu$ and the diffusion
 tensor $D^{\mu\nu}$ are given by
 \be
 A^\mu=\int f_*\Delta p^\mu_*\,F \sigma d\Omega\sqrt{g} {d^3p_*\over p_{*0}},
 \ee{48}
 \be
 D^{\mu\nu}={1\over2}\int f_*\Delta p^\mu_*\Delta p^\nu_*\,F \sigma d\Omega\sqrt{g} {d^3p_*\over
 p_{*0}}.
 \ee{49}

In Sec. III we made an analogy with Brownian motion in the relativistic case, that analogy was made through
scattering assumptions based on specific properties like mass and density ratios of the constituents of the
system. These assumptions allow us to made a physical analogy with Brownian motion and none of the mathematical
stochastic properties of Brownian motion were used. In fact the evolution in this version of Brownian motion is
provided by the collisions while in other works \cite{RZ}, \cite{Debb1}, \cite{Debb2}, \cite{Hang1} and
\cite{Hang2} is given by a stochastic term.

Recently in \cite{Hang4} is shown that a modified Maxwell-J\"{u}ttner distribution could be obtained from a
modified principle of maximum entropy that take into account the properties of Lorentz group. This is a hint
that the relativistic Boltzmann equation might require a slight modification. Indeed the relativistic kinetic
theory considered here is not the only version, there are others proposals like
\cite{Schive} and \cite{GC} which try to improve some points of the theory.

\subsection*{ACKNOWLEDGMENTS} One of the authors (G.C.A.) is very grateful with H. A. Morales-T\'ecotl and
L. Dagdug for valuable discussions and advices which help him to develop this work. He also thanks R. M. Velasco
for useful comments as well as the hospitality of the UFPR during the 2006 visit when this work was done. This
work was partially supported by a scholarship from CONACyT (G.C.A.) and by CNPq (G.M.K.).

\end{document}